\documentclass{elsart}
\usepackage{natbib}

\usepackage{epsfig}
\usepackage{amsmath}
\usepackage{amssymb}

\newcommand{\eg}{{\em e.g. }}

\newcommand{\gsim}{\gtrsim}
\newcommand{\lsim}{\lesssim}

\newcommand{\de}{\delta}
\newcommand{\De}{\Delta}
\newcommand{\ep}{\epsilon}
\newcommand{\ga}{\gamma}

\newcommand{\la}{\lambda}
\newcommand{\Om}{\Omega}

\newcommand{\be}{\begin{equation}}
\newcommand{\ee}{\end{equation}}
\newcommand{\bee}{\begin{equation*}}
\newcommand{\eee}{\end{equation*}}
\newcommand{\bea}{\begin{eqnarray}}
\newcommand{\eea}{\end{eqnarray}}
\newcommand{\bean}{\begin{eqnarray*}}
\newcommand{\eean}{\end{eqnarray*}}

\newcommand{\bk}{{\mathbf k}}

\newcommand{\bB}{{\mathbf B}}

\newcommand{\bn}{{\bf n}}

\begin{document}
\runauthor{Ruth Durrer}
\begin{frontmatter}
\title{Cosmic Magnetic Fields and the CMB}
\author{Ruth Durrer}
\address{Universit\'e de Gen\`eve, Ecole de Physique}
\address{24 Quai E. Ansermet, CH-1211 Gen\`eve}
\address{Switzerland}

\begin{abstract}
I describe the imprint of primordial magnetic fields on
the CMB. I show that these are observable only if the field
amplitude is of the order of $B\gsim 10^{-9}G$ on Mpc scale. I
further argue that such fields are strongly constrained by the
stochastic background of gravity waves which they produce.
Primordial magnetic fields, which are strong enough to be seen in
the CMB, are compatible with the nucleosynthesis bound, only if
their spectrum is close to scale invariant, or maybe if helical
magnetic fields provoke an inverse cascade.
For helical fields, the CMB signature is
especially interesting. It contains parity violating T-B and E--B
correlations.
\end{abstract}
\end{frontmatter}

\section{Introduction}
In observational cosmology we try to constrain the history of the
Universe by the observation of relics. The best example of this is
the cosmic microwave background (CMB) which represents not only a
relic of the time of recombination, $t \simeq 3.8\times 10^5$ years after
the big bang, but probably also of a much earlier epoch, $t\sim
10^{-35}$sec, when inflation took place. Another such relic is the
abundance of light elements established during primordial nucleosynthesis
at $t\simeq 100$ sec.

But there are other very interesting events which might have left
observable traces, relics, in the universe. Most notably
confinement at $t \simeq10^{-4}$ sec or the electroweak transition
at $t \simeq 10^{-10}$sec which may have generated the observed
baryon asymmetry in the Universe, see Ref.~\cite{barass}. It has been
proposed that confinement and, especially the electroweak phase
transition might also lead to the formation of primordial magnetic
fields which then can seed the magnetic fields observed in
galaxies and clusters, see Refs.~\cite{mfew,mfconf}.

Magnetic fields are ubiquitous in the Universe. Most galaxies, like
the milky way, are permeated by a magnetic field of the order of a
few $\mu$Gauss, see Ref.~\cite{obsgal}. But also clusters of galaxies have
magnetic fields of the same order, see Ref.~\cite{obsclu}.

If these fields are due to the amplification of seed
fields by to the contraction of the cosmic plasma during the
process of galaxy formation, seed fields of the order of
$10^{-9}$G are needed. However if they are amplified via a non
linear dynamo mechanism, seed fields as little as about $10^{-22}$
Gauss might be sufficient, see Ref.~\cite{Davis}. It is not clear
whether seed fields are really needed. It may be possible that charge
separation processes during structure formation can lead to
currents which generate magnetic fields without the presence of any
seed fields. It is still a matter
of debate whether second order perturbations alone can induce sufficient
charge separation, and therefore currents, to provoke the formation the observed
fields, see Refs.~\cite{Riotto05,Ichiki05}.

In this {\em talk} I assume that primordial magnetic fields have been generated
at early time with some initial spectrum, and I discuss their effects on the CMB.
Then I derive the spectrum of 'causal' magnetic fields (i.e. fields
generated during a non-inflationary phase of the universe).
We shall see that magnetic fields, especially causal magnetic
fields, are very strongly constrained by the gravity wave background
which they induce.
I shall finally indicate some possible ways out of this stringent constraints.

\section{Effects of magnetic fields on the CMB}

\subsection{A constant magnetic field}
A spatially constant magnetic field affects the geometry of the universe by introducing shear.
It generates an anisotropic stress, $\Pi_{ij}\propto B_iB_j -\frac{1}{3}B^2\de_{ij}$.
This leads to a well studied (anisotropic) homogeneous model, the so called
Bianchi VII model, see Ref.~\cite{Bianchi}. The propagation of CMB photons from the last scattering
surface into our antennas through the Bianchi~VII geometry leads to anisotropies.
Comparing these with the observed anisotropies, we obtain limits on a constant
magnetic field. Comparing, e.g., the CMB quadrupole with the one induced by a
constant magnetic field, we can limit the field amplitude to
$B< 6.8\times 10^{-9}(\Om_mh^2)^{1/2}$Gauss, see Ref.~\cite{Barrow97}.

\begin{figure}
\begin{center}
\begin{minipage}{1\linewidth}
\centering
\epsfig{figure=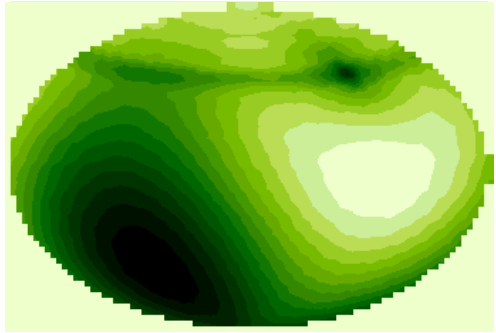,width=6.5cm}
\caption{The CMB anisotropy pattern from a constant magnetic field,
from Ref.~\cite{durrer98}. \label{mag1}}
\end{minipage}
\end{center}
 \end{figure}

As usual, we decompose the CMB temperature fluctuations into spherical harmonics,
$$ \frac{\De T}{T}(\bn) = \sum_{\ell,m} a_{\ell m}Y_{\ell m}(\bn) ~, $$
where $\bn$ denotes the direction of observation. The coefficients $a_{\ell m}$
can be considered as the amplitude of the spin $\ell$ contribution to the
temperature fluctuation (with $z$-component $m$). In a statistically isotropic universe
$\langle a_{\ell m}a^*_{\ell' m'}\rangle = C_\ell\de_{\ell\ell'}\de_{mm'}$.
Different $\ell$s and $m$s are uncorrelated.
Since the presence of  a constant magnetic field breaks statistical isotropy,
see Fig.~\ref{mag1}, it leads to correlations of $a_{\ell m}$'s with
different values of $\ell$. A detailed analysis shows that, see 
Ref.~\cite{durrer98}, for a constant magnetic field in $z$-direction,
there are non-vanishing correlators with $\ell\neq \ell'$, namely
$\langle a_{\ell-1 m}a^*_{\ell+1 m}\rangle \neq 0$. The magnetic field energy momentum tensor
acts like a spin-2 field and leads to transitions from $\ell-1$ to $\ell +1$ thereby
correlating these amplitudes.

Limiting such off-diagonal correlations with the COBE data also leads to bounds of the
order of $B< 3\times 10^{-9}$Gauss, see Ref.~\cite{durrer98}.

It is not surprising that the limits from the quadrupole and from the off--diagonal
correlators are comparable, since  $\Om_B = 10^{-5}\Om_{rad} (B/10^{-8}$Gauss$)^2$.
Therefore magnetic fields of the order $3\times 10^{-9}$Gauss will leave of order
10\% effects on the CMB anisotropies while $10^{-9}$Gauss will typically contribute
 1\% effects. It is thus clear that we can never detect magnetic fields of the order
 of $10^{-22}$ Gauss with CMB observations.

\subsection{CMB anisotropies from stochastic magnetic fields}
To obtain thelimits for a constant magnetic field given in the
previous section, we have only taken into account that the energy
momentum tensor of the magnetic field affects 
the geometry. When the CMB photons then propagate along the correspondingly
modified geodesics, the CMB sky becomes anisotropic. These anisotropies are to be added
to the anisotropies due to the usual inflationary perturbations.
But there are also other effects of magnetic fields on CMB photons. In this section I
briefly describe them all.

The magnetic field energy momentum tensor contains scalar, vector and tensor components
which all modify the spacetime geometry and therefore affect the propagation of photons
as we have discussed above.
For tensor perturbations this is all there is, see Ref.~\cite{durrer00}.

In addition, via its coupling to the charged electron--proton plasma,
the magnetic field  generates vector perturbations in the plasma
velocity which oscillate, so called Alfv\'en waves, see
Refs.~\cite{Jackson,durrer98}. This is a new phenomenon. Purely 
gravitational vector perturbation do not show oscillatory behavior. If
we would be able to constrain (or better even detect!) possible vector
contributions to the CMB anisotropies this may provide very important
limits on primordial magnetic fields due to the specific signal 
from Alfv\'en waves.

\begin{figure}
\begin{center}
\begin{minipage}{1\linewidth}
\centering
\epsfig{figure=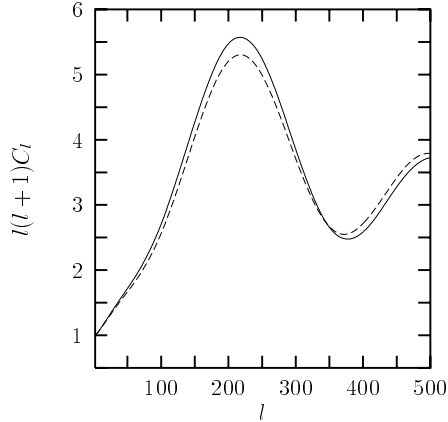,width=7.5cm}
\vspace*{-3cm}\\
\caption{The modification of the CMB anisotropy spectrum due to fast magneto-sonic waves,
for a magnetic field amplitude of $3\times 10^{-7}$Gauss. From Ref.~\cite{Adams96}.
\label{sonic}}
\end{minipage}
\end{center}
 \end{figure}

There are also two types of scalar waves in the presence of magnetic
fields, the so called fast and slow magneto-sonic waves. They are induced
by the scalar perturbations  of the magnetic field and the charged plasma,
see Ref.~\cite{Adams96}.
Fast magneto-sonic waves are simply the ordinary sound waves which are modified
due to the presence of the magnetic fields, they acquire a somewhat higher
sound speed     $c_s^2 \rightarrow c_s^2 + (\bk\cdot\bB)^2/(4\pi\rho)$
leading to a slight shift of the acoustic peaks which might be detectable,
see Fig.~\ref{sonic}.
Slow magneto-sonic waves provide a new form of waves due to the interaction
of the charged plasma with the magnetic field, see
Refs.~\cite{Jackson,Adams96}. They have a very low sound speed and
their effect on the CMB is small.

\subsection{Limiting magnetic fields with the CMB}
To formulate limits on a stochastic primordial magnetic field
distribution, we have to define an appropriate way to describe the
latter. For this we define the magnetic field spectrum in the form
\be\label{Bspec}
a^4(t)\langle B_i(\bk)B^*_j(\bk')\rangle k^3 =\left\{ \begin{array}{lll}
\frac{1}{2}\de^3(\bk-\bk')
\left(\de_{ij} -\hat k_i\hat k_j\right)(k\la)^{n+3}B_\la^2 & \mbox{for} & k< k_D\\
 0  & \mbox{for} & k> k_D~. \end{array}\right.
\ee
Here $\la$ is some arbitrary length scale, usually the one of
interest in a given problem. The average field amplitude at some scale
$k^{-1}$ is
$$B^2\!\mid_{k^{-1}} \simeq \bB(k)^2k^3 =(k\la)^{n+3}B_\la^2~,$$
so that $B_\la$ is simply the amplitude of the magnetic field at
that scale $k^{-1}=\la$. We normalize the scale factor $a(t)$ to
today, $a(t_0)=1$. Note that we have to use the projector $\de_{ij}
-\hat k_i\hat k_j$, onto the plane normal to $\bk$ so that the
Maxwell equation $\nabla\cdot\bB=0$ is verified, $\hat\bk$ denotes
the unit vector in direction $\bk$. The scale $\la_D=k_D^{-1}$ is
the damping scale below which magnetic fields are converted into
heat usually be fluid viscosity. This scale depends on time and
has to be computed using the magneto-hydrodynamic equations and
determining the viscosity of the different components of the
cosmic plasma, see Refs.~\cite{viscous,Caprini02}. The scale $\la_D$ is
steadily growing as the cosmic plasma dilutes, it amounts to
several parsecs at present time. The power $n$ is the spectral
index of the magnetic field spectrum, $n=-3$ corresponds to a
scale invariant spectrum. In order for the magnetic field not to
diverge on large scales, we must require $n\ge -3$.

\begin{figure}
\begin{center}
\begin{minipage}{1\linewidth}
\centering \epsfig{figure=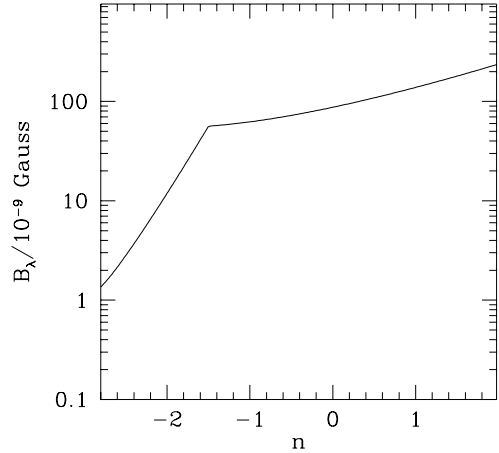,width=7.5cm} \caption{Limits
on stochastic magnetic fields from the tensor anisotropies induced
in the CMB are shown as a function of the spectral index. The
characteristic scale chosen is $\la=0.1h^{-1}$Mpc. From
Ref.~\cite{durrer00}. \label{fig:limit}}
\end{minipage}
\end{center}
 \end{figure}

In Fig.~\ref{fig:limit} we show limits on a stochastic magnetic
field which have been derived taking into account only tensor
fluctuations generated in the CMB,  and requiring that they do not
overproduce the CMB anisotropies, see Ref.~\cite{durrer00}.

\begin{figure}
\begin{center}
\begin{minipage}{1\linewidth}
\vspace{1cm}
\centering \epsfig{figure=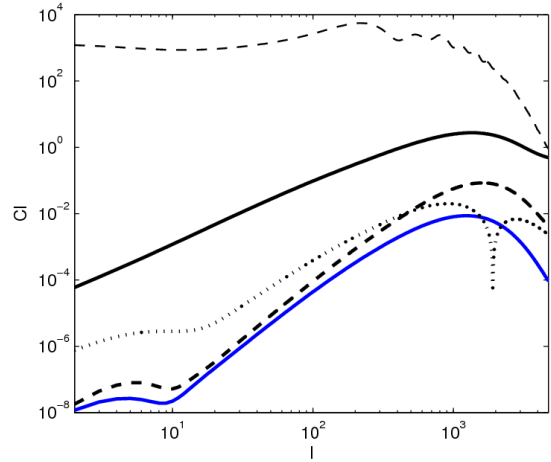,width=7.5cm} \caption{The
$C_\ell$s from vector temperature anisotropies and polarization
induced by a stochastic magnetic field with a nearly scale invariant
spectrum, $n=-2.99$ and an amplitude $B_\la = 3\times
10^{-9}$Gauss are shown (again $\la=0.1h^{-1}$Mpc is chosen).
The top curve shows the usual scalar TT
spectrum. The curves below show the vector perturbations induced by the
magnetic field: the next (solid) curve is the TT spectrum, the
dotted curve is the TE spectrum, the dashed curve the
B-polarization and the lowest solid curve (blue) the
E-polarization spectrum. From Ref.~\cite{Lewis04}. \label{lewis}}
\end{minipage}
\end{center}
\end{figure}

 Taking into account all other effects on CMB anisotropies similar
 limits have been obtained, see Refs.~\cite{Mack02,Lewis04}. The vector mode
 fluctuations, not taking into account Faraday rotation and a
 helical component (discussed in the next subsections) are shown
 in Fig.~\ref{lewis}.

\subsection{Polarization}
Polarization is affected by the presence of magnetic fields mainly
by two mechanisms. First, the gravitational field which is
modified by the magnetic field energy momentum tensor leads to a
change in the evolution of polarization (which is parallel
transported along the photon geodesics). This is taken into account if
Fig.~\ref{lewis}. The second effect is
Faraday rotation: the magnetic field polarizes the electrons in
the plasma, which leads to a rotation of the polarization of CMB
photons scattering with them. This can rotate E-polarization into
B-polarization! The effect is, however, frequency dependent $
\propto 1/\nu^2$ and can therefore, in principle, be distinguished
from intrinsic B-polarization. At $\ell \sim 1000$, Faraday rotation
induced a B-polarization from the ordinary scalar E-polarization
of roughly $10^{-3}\mu$K$($B$/10^{-9}$G$)(\nu/10$GHz$)^{-2}$,
see \eg Ref.~\cite{Takada01}. The modification of polarization
due to Faraday rotation in the presence on a magnetic field has
been discussed in detail in Ref.~\cite{TinaFar}.

\subsection{Helical magnetic fields}
The magnetic field spectrum which we have given in Eq.~(\ref{Bspec})
is the most general power law spectrum which is invariant under
parity. If we allow for parity violation, we can add another
term,
\be\label{Bspec2}
a^4\langle B_i(\bk)B^*_j(\bk') =\left\{ \begin{array}{lll}
\de^3(\bk-\bk')
\left( (\de_{ij} -\hat k_i\hat k_j)S(k) +i\ep_{ijm}\hat k_m A(k)\right) &
\mbox{for} & k< k_D\\
 0  & \mbox{for} & k> k_D~. \end{array}\right.
\ee
Here $\ep_{ijm}$ is the totally antisymmetric tensor in three dimensions.
This is the most generic expression for the correlation function of a
magnetic field distribution which is stochastically homogeneous and isotropic.
$S$ and $A$ are functions of the modulus of $\bk$, in the simplest
case they are pure power laws with index $n_S$ and $n_A$.
The second term is parity odd and can only
be generated by parity violating interactions (e.g. at the
electroweak phase transition, see Ref.~\cite{Vachaspati01} ).
 In the CMB such a
term induces parity odd correlations between temperature
anisotropies and B-polarization and between E- and B-polarization
which have been calculated in Ref.~\cite{Caprini04}, see Fig.~\ref{helic}
below. The results are expressed in terms of
the density parameter of the parity even and odd magnetic field contributions,
$\Om_S$ and $\Om_A$. Since the parity odd part actually contributes negatively
to the energy density, $\Om_B = \Om_S -\Om_A$, we have to require
$\Om_S>\Om_A$. For a pure power law spectrum, positivity requires $n_A>n_S$.
\begin{figure}
\begin{center}
\begin{minipage}{1\linewidth}
\centering \epsfig{figure=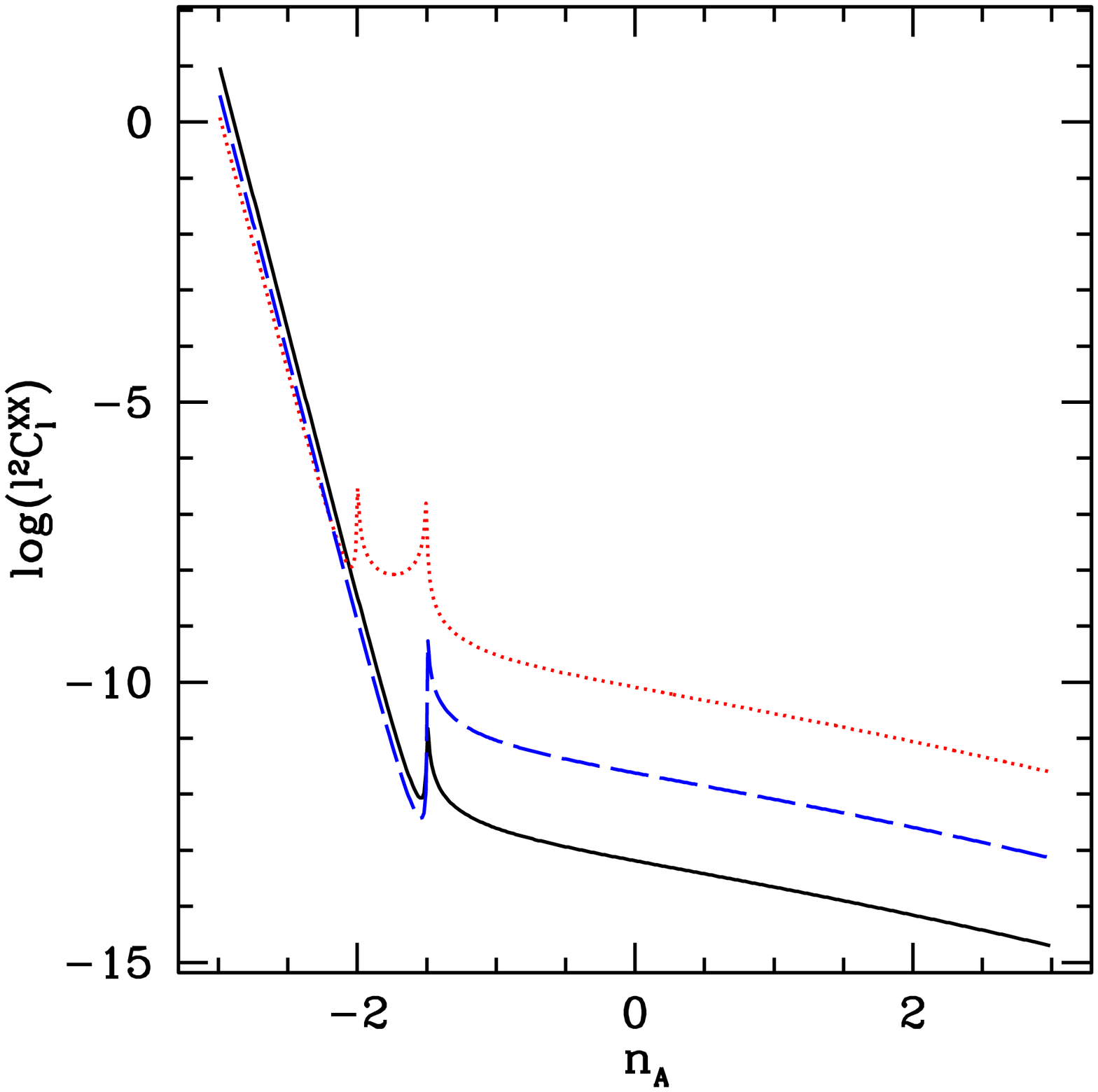,width=6.5cm} \epsfig{figure=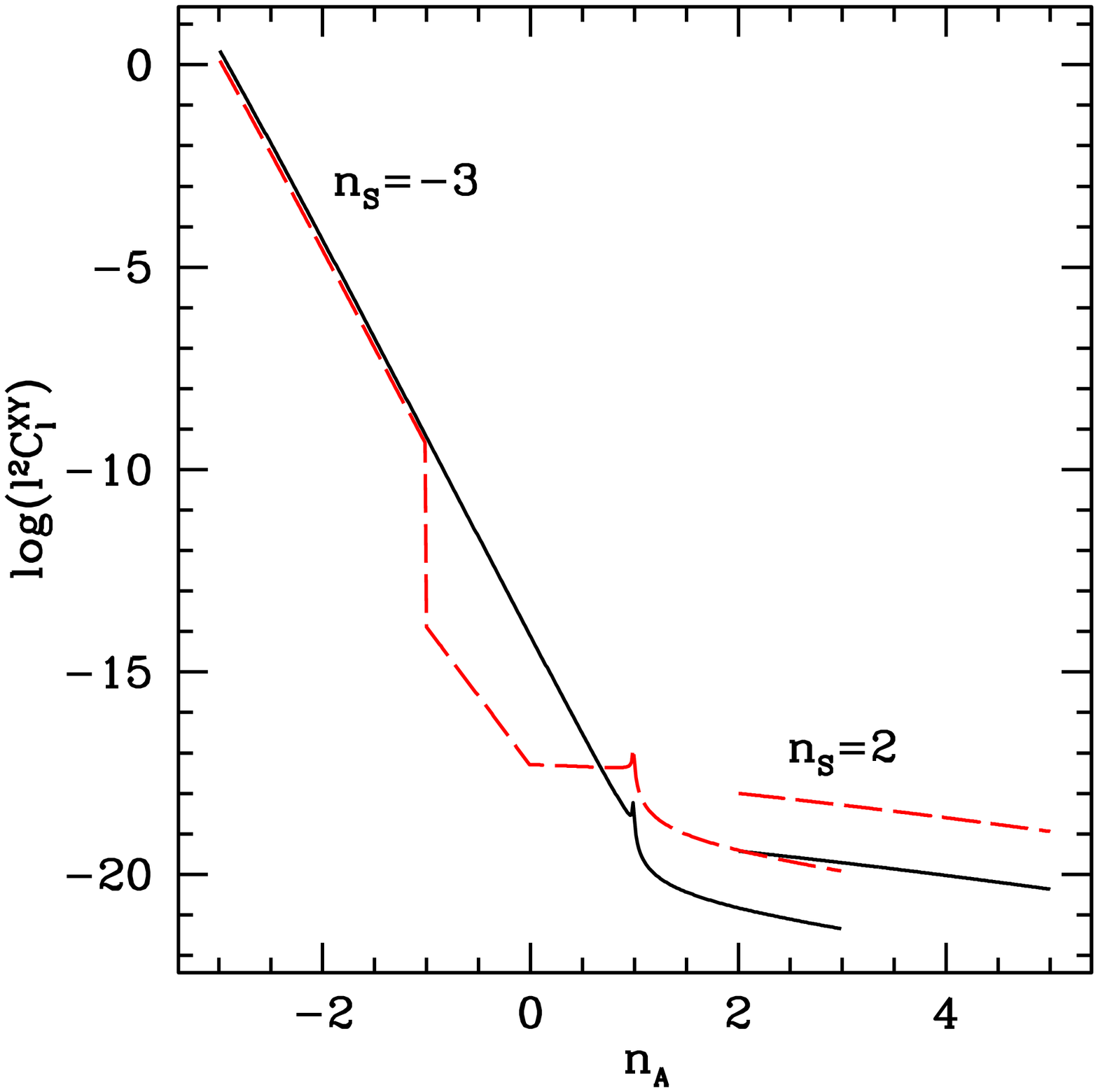,width=6.5cm}
\caption{Left: The temperature anisotropy (solid, black), the E-polarization (dotted, red)
and the T--E-correlation (dashed blue) induced from the parity violating contribution are
shown as functions of $n_A$ for $n_S= - 3$, in units of
$(\Om_A\Om_S/\Om_{\mathrm rad}^2)\log^2(z_*/z_{eq})$. Here $z_*$ is the redshift
at which the magnetic field is generated (supposed to be before equality)
and $z_{\mathrm eq}$ is the redshift of matter and radiation equality.
The B-polarization is equal to the E-polarization within the semi-analytic
approximation used for this results.\newline
Right: The T--B (solid, black), and E--B (dashed, red) correlators are shown in the same units.
\newline
The peaks and kinks in the curves are not physical. They are due to
the breakdown of 
semi-analytic approximations used in the calculations.
 From Ref.~\cite{Caprini04}. \label{helic}}
\end{minipage}
\end{center}
 \end{figure}

\section{Causality and the magnetic field spectrum}
If magnetic fields have been produced at some time $\eta_*$  when the universe
was not inflating, the correlations $\langle B(x)B(y)\rangle$ vanish
for sufficiently large distances, e.g $|x-y|> \eta_*$. Hence the
correlation function is a function 
with compact support and therefore its Fourier transform is analytic.
For the spectrum given above~(\ref{Bspec2}) this means that, at 
sufficiently small
values of $k$, the functions $S$ and $A$ are dominated by one power,
$S(k) = S_0k^{n_S}$ and $A(k) = A_0k^{n_A}$. Analyticity then implies
that $n_S\ge 2$ is an even integer,
and $n_A\ge 1$ is an odd integer. Positivity of the magnetic field energy requires
in addition $n_A> n_S$, hence $n_A \ge 3$, see Ref.~\cite{Caprini03}.
Causal magnetic field spectra are therefore very blue and might lead to better
constraints on small scales than on large scales.
As we shall explain in the next section, where we study not the effects on the CMB
but simply the production of gravity waves by stochastic magnetic fields, this is
indeed the case.

\section{Limiting primordial magnetic fields with gravity waves}
The fact that causal magnetic fields have very blue spectra, suggests
that the best 
limits for them can be obtained on small scales. The energy density of
gravity waves produced from stochastic magnetic fields with a given spectral index $n_S$
and amplitude $B_\la$  is always
dominated by the contribution from small wavelengths.

At horizon crossing the magnetic fields convert a sizable fraction of their energy
into gravity waves. This gravity wave background is not damped by
subsequent interactions with the cosmic plasma, it is simply diluted by the expansion
of the universe like any other radiation component.
Comparing the produced intensity of gravity waves with the nucleosynthesis bound
leads to very stringent limits on the causal production of cosmic magnetic
fields, see Refs.~\cite{Caprini02,Caprini05}.

\subsection{Limits for non-helical fields}
\begin{figure}
\begin{center}
\begin{minipage}{1\linewidth}
\centering \epsfig{figure=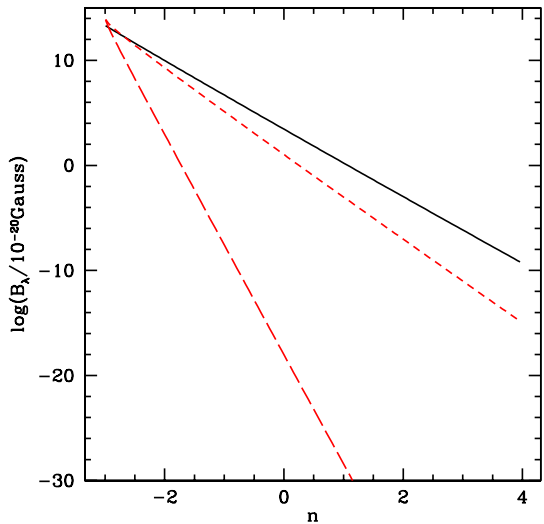,width=7.5cm} \caption{Limits
on stochastic magnetic fields from the induced stochastic gravity wave background
as a function of the spectral index $n$. Only the symmetric (parity invariant) contribution
has been considered. The solid (black) line is the limit implied from the magnetic field
energy density itself. The short-dashed line represents the limit from the gravity
wave background for magnetic fields produced at the electroweak phase transition while
the long-dashed line limits magnetic fields form inflation.
$B_\la$ is normalized to today at the scale of $\la=0.1$Mpc.
 From Ref.~\cite{Caprini02}. \label{fig:BlimGW}}
\end{minipage}
\end{center}
\end{figure}

In Ref.~\cite{Caprini02} the limit on magnetic fields from the
fact that the gravity wave background they produce should be below
the nucleosynthesis limit, $\Om_{GW} \le 0.1\times\Om_{\mathrm rad}$
is derived, see Fig.~\ref{fig:BlimGW}. The energy density in gravity
waves is entirely dominated
by the short wavelength contributions and therefore by the cutoff $k_D(\eta_*)$ which
is conservatively set to $k_D(\eta_*) = \eta_*^{-1}$. A recent calculation with a more
realistic value for $k_D(\eta_*)$ has even somewhat improved the limits,
see Ref.~\cite{Caprini06}.
It is interesting to note that the gravity wave limit from the nucleosynthesis bound
is stronger than the limit coming from the magnetic field energy density itself (solid
black line in Fig.~\ref{fig:BlimGW}). This comes from the fact that for the examples
depicted, the damping scale at nucleosynthesis, which represents the
upper cutoff for the magnetic field energy spectrum, is larger
than the horizon scale at the time of formation of the magnetic field,
$\eta_*$, which is the upper cutoff for
the gravity wave energy. Therefore, the magnetic field energy on this scale
which has not been converted in gravity waves has been converted into
heat by the time of nucleosynthesis.

\subsection{Ways out}
 The limits presented in the previous subsection are extremely stringent. They
 suggest that, for example magnetic fields which are causally generated at
 the electroweak phase transition cannot lead to fields larger than $10^{-28}$Gauss
 at the scale of 0.1Mpc, which is insufficient for amplification to the
 presently observed fields even with a strong dynamo mechanism.
 To arrive at this result, we have used that apart from being damped at small scales,
 $k>k_D(\eta)$ the magnetic field simply follows the expansion
 of the cosmic plasma on large scales, so that $B\propto 1/a^2$.
 'Normal' magnetic fields cannot do better. Their spectra evolve by damping on
 small scales and cascade (moving power from larger into smaller scales) on large,
 but sub-horizon scales.
 However, analytical arguments and numerical simulations show that {\em helical}
 magnetic fields can actually invoke an {\em inverse cascade}: they can transport power
 from smaller into larger scales. A bit like a cosmic string network,
 magnetic flux lines
 which intersect can reconnect and produce larger scale coherence, see
 Refs.~\cite{Vachaspati01,Jedamzik02, Hindmarsh03, Brandenburg05}.
 A trustable, quantitative evaluation of this inverse cascade is still missing.
 But it may represent a way out of the stringent limits presented above.

 For inflation the situation is more delicate. 
If the inflaton isa pseudo-scalar field it can in principle couple to
the electromagnetic field in a ways that violates parity (a term of
the form $\phi F_{\mu\nu}*F^{\mu\nu}$ in the Lagrangian). This does
indeed lead to a helical magnetic field. This has been studied in
Ref.~\cite{carroll} and it has been found that the amplitude
is much too small to be relevant~\cite{carroll}.  
However, the causality requirement drops at inflation and correlations
can be generated on arbitrary large scales. 
Therefore, the spectral index is not limited to 2 by causality, but we could in
 principle have a spectral index as small as $n\simeq -3$. As it is show in
 Fig.~\ref{fig:BlimGW}, for this spectral index the magnetic field
 limit raises to  $B_\la \lsim 10^{-9}$Gauss, a field which is large
 enough to lead to the observed fields in galaxies and clusters by
 simple adiabatic contraction and which might even be detected in the
 CMB.\\ 
However, since the electromagnetic field is conformally coupled, it is not
produced during 'ordinary' inflation. It is, however produced \eg
during a pre-big bang phase with a dynamical dilaton, 
see Ref.~\cite{Gasparini95}. There, large scale coherent electromagnetic
fluctuations are generated. Due to the high conductivity of the cosmic plasma,
the electric field is rapidly dissipated and it remains a magnetic field.
Also non-standard couplings of the electromagnetic field during inflation
can lead to the generation of magnetic fields.\\
Typically one finds spectra with $n \simeq 0$,
see Ref.~\cite{TurnerWidrow88} for which the gravity wave constraints
shown in Fig~\ref{fig:BlimGW} are again far too stringent. But also
spectra with $n \simeq -3$ have been proposed for some very specific
(albeit not very well motivated) situation, see~\cite{Bamba04}.

\section{Conclusions}
In this talk I have described the physical effects by which
primordial magnetic fields leave an imprint on the CMB.
Since $\Om_B = 10^{-5}\Om_\ga(B/10^{-8}$G$)^2$, this is only detectable if
$B\gsim 10^{-9}$G on CMB scales. If $n>-3$, this means that the
magnetic fields on smaller scales are larger. Then they are usually strongly
constrained by  the induced gravity wave background.

 To generate the observed galactic or cluster magnetic fields by
 simple contraction, 
 seed fields of the order of $B\simeq 10^{-9}$G on about 1Mpc scale are needed.
 Dynamo amplification requires seed fields of about $10^{-28}$G.

 The induced gravity wave background limits causally produced
(non-helical) fields from the electroweak phase transition ($n=2$)to
 $B< 10^{-30}$G on 1Mpc scale and fields from inflation with spectral index
$n\simeq 0$ to $B<10^{-43}$G. Only scale invariant magnetic seed
 fields, $n=-3$, may be as large as $10^{-9}$G and therefore  leave a
 detectable imprint on the CMB.  
 Helical fields induce an inverse cascade leading to larger fields on
 large scales. Therefore, the above limits do not apply for them.

 Recently it has also been argued that currents induced by charge separation,
 in 2nd order cosmological perturbation theory, may
 generate seed fields at much later times, after recombination, which are not constrained by
 the nucleosynthesis bound. The true amplitude of the fields obtained in this way is still
 a matter of debate and varies between $(10^{-23}$ --- $10^{-16})$G, see
 Refs.~\cite{Riotto05,Ichiki05}. This interesting possibility
 certainly deserves more work
 (see also Ref.~\cite{durrerScience}).
 \vspace{0.5cm}

{\large\bf Acknowledgment}\\
 I thank the organizers of this memorial meeting for the wonderful atmosphere
 at which this workshop took place, and
 Francesco Melchiorri for the stimulating discussions I had with him when he was
 still among us. I miss him.

\end{document}